\documentclass[aps, showpacs, showkeys,nofootinbib,floatfix]{revtex4}

\usepackage{amssymb}
\usepackage{amsmath}
\usepackage{graphicx}

\begin{document}

\title{Cosmography: Supernovae Union2, Baryon Acoustic Oscillation, Observational Hubble Data and Gamma Ray Bursts}

\author{Lixin Xu\footnote{Corresponding author}}
\email{lxxu@dlut.edu.cn}

\affiliation{Institute of Theoretical Physics, School of Physics \&
Optoelectronic Technology, Dalian University of Technology, Dalian,
116024, P. R. China}

\affiliation{College of Advanced Science \& Technology, Dalian
University of Technology, Dalian, 116024, P. R. China}

\affiliation{Korea Astronomy and Space Science Institute,
Yuseong Daedeokdaero 776,
Daejeon 305-348,
R. Korea}

\author{Yuting Wang}

\affiliation{Institute of Theoretical Physics, School of Physics \&
Optoelectronic Technology, Dalian University of Technology, Dalian,
116024, P. R. China}

\begin{abstract}
In this paper, a parametrization describing the kinematical state of
the universe via cosmographic approach is considered, where the
minimum input is the assumption of the cosmological principle, i.e.
the Friedmann-Robertson-Walker metric. A distinguished feature is
that the result does not depend on any gravity theory and dark
energy models. As a result, a series of cosmographic parameters
(deceleration parameter $q_0$, jerk parameter $j_0$ and snap
parameter $s_0$) are constrained from the cosmic observations which
include type Ia supernovae (SN) Union2, the Baryon Acoustic
Oscillation (BAO), the observational Hubble data (OHD), the high
redshift Gamma ray bursts (GRBs). By using Markov Chain Monte Carlo
(MCMC) method, we find the best fit values of cosmographic
parameters in $1\sigma$ regions: $H_0=74.299^{+4.932}_{-4.287}$,
$q_0=-0.386^{+0.655}_{-0.618}$, $j_0=-4.925^{+6.658}_{-7.297}$ and
$s_0=-26.404^{+20.964}_{-9.097}$ which are improved remarkably. The
values of $q_0$ and $j_0$ are consistent with flat $\Lambda$CDM
model in $1\sigma$ region. But the value of $s_0$ of flat
$\Lambda$CDM model will go beyond the $1\sigma$ region.
\end{abstract}


 \hfill TP-DUT/2011-01

\maketitle

\section{Introduction}

The kinematical approach to describe the status of universe is
interesting for its distinguished feature that it does not rely on
any dynamical gravity theory and dark energy models. Then it becomes
crucial for its potential ability to distinguish cosmological models
when a flood of dark energy models and modified gravity theories are
proposed to explain the current accelerated expansion of our
universe. This late time accelerated expansion of our universe was
firstly revealed by two teams' observation of type Ia supernovae
\cite{ref:Riess98,ref:Perlmuter99}. In general, via the Taylor
expansion of the scale factor $a(t)$ in terms of cosmic time $t$,
the dimensionless coefficients $q_0$, $j_0$ and $s_0$ named
deceleration, jerk and snap parameters are defined respectively, for
the detailed forms please see Eq. (\ref{eq:q_0}, \ref{eq:j_0},
\ref{eq:s_0}) in the following. For convenience, they are dubbed as
{\it cosmographic parameters}. These cosmographic parameters, which
current values can be determined by cosmic observations, describe
the kinematical status of our universe. For example, the present
value of Hubble parameter $H_0$ describes the present expansion rate
of our universe, and a negative value of $q_0$ means that our
universe is undergoing an accelerated expansion. This kind of
approach is also called cosmography
\cite{ref:cosmography,ref:ST2006}, cosmokinetics
\cite{ref:cosmokinetics,ref:cosmopara}, or Friedmannless cosmology
\cite{ref:Friedmannless1,ref:Friedmannless2}. Recently, this
approach was considered by using SN in Ref. \cite{ref:kinematic2},
SN+GRBs in Ref. \cite{ref:highcosmography} and SN+OHD+BAO in
\cite{ref:kinematicXu}, where the current status of our universe can
be read. On the other hand, for a concrete dark energy model or
gravity theory, when the Friedmann equation is arrived the
corresponding cosmographic parameters can be derived by simple
calculation. As a consequence, the corresponding parameter spaces
can be fixed from cosmographic parameters space without implementing
annoying data fitting procedure. However, the reliability of the
cosmographic approach depends crucially on how the cosmographic
parameter space is shrunk, in other words, the improvement of the
figure of merit (FoM). That is the main motivation of this paper. In
general, when more cosmic observational data sets are added to
constrain model parameter space, the more degeneracies between model
parameters will be broken. Also the FoM will be improved. So, to
investigate the current status of our universe and to improve the
FoM, the cosmographic parameters will be determined by more cosmic
observations. When the SN and GRBs are used as distance indicators,
the Hubble parameter $H_0$ and the absolute magnitudes of SN and
GRBs are treated as notorious parameters and marginalized. That is
to say, SN and GRBs can not fix the current value of Hubble
parameter $H_0$. That is what the authors have done in Ref.
\cite{ref:kinematic2,ref:highcosmography} where the cosmographic
parameters $q_0$, $j_0$ and $s_0$ were investigated. However, the
cosmographic parameters permeate in a relative larger space. Of
course, to describe the kinematical status of our universe well, one
has to shrink the parameter space efficiently. Fortunately, when the
Hubble parameter $H_0$ is fixed as done in Ref.
\cite{ref:kinematicXu}, the parameter space is pinned down
effectively. When the snap parameter $s_0$ is included, high
redshift observations should be added. So, in this paper we are
going to use SN, BAO, GRBs, OHD to investigate the cosmographic
approach. When SN data sets are used, the systematic errors are
included. The BAO are detected in the clustering of the combined
2dFGRS and SDSS main galaxy samples, so it is helpful to break the
degeneracies between parameters. The OHD data sets are used to fix
the Hubble parameter $H_0$. Higher redshift data ponits are from GRBs where the correlation parameters are calibrated via cosmographic approach synchronously. For the detailed description of these
data sets, please see the Appendix \ref{app:obervations}.

This paper is structured as follows. In Section \ref{sec:cg}, the definition of cosmographic parameters and basic expansions with respect to redshift $z$ are presented, where to consider the convergence issue, the map from $z\in (0,\infty)$ to $y=z/(1+z) \in(0,1)$ is adopted. To the expansion truncation problem, we compare the expansions 
with $\Lambda$CDM model in the range of redshift involved in this paper. The relative departure of Hubble parameter from that of $\Lambda$CDM model is up to $20\%$ at the redshift $z\sim 1.75$.
The dfference of distance modulus between the expansion of luminosity distance and that of $\Lambda$CDM model is less than $1.6$. Section \ref{sec:results} are the main results of this paper. To obtain these results, the cosmic observational data sets from SN Ia, BAO, OHD and GRBs and MCMC method are used. The detailed descriptions are shown in the Appendix \ref{app:obervations}. The main points of this paper are listed as follows: 1). BAO and OHD are used to shrink the model parameter space \footnote{After our work, the papers used BAO and OHD appeared in arXiv: J. Q. Xia, et. al, arXiv:1103.0378 and S. Capozziello, et. al, arXiv:1104.3096}. 2). The calibration of GRBs and constraint to cosmographic parameters are carried out synchronously. In this way the so-called circular problem is removed. We summarize the results in Tab. \ref{tab:results} and Fig. \ref{fig:cases} and Fig. \ref{fig:hubmod}. Section \ref{sec:con} is a brief conclusion. 

\section{Cosmographic Parameters} \label{sec:cg}

The minimum input of the cosmographic approach is the assumption of
the cosmological principle, i.e. the Friedmann-Robertson-Walker
(FRW) metric
\begin{equation}
ds^2=-dt^2+a^2(t)\left[\frac{dr^2}{1-kr^2}+r^2(d{\theta}^2+\sin^2{\theta}d{\phi}^2)\right],
\end{equation}
where the parameter $k=1,0,-1$ denotes spatial curvature for closed,
flat and open geometries respectively. In this paper, we only
consider the spatially flat case $k=0$.

\begin{figure}[!htbp]
\includegraphics[width=8.2cm]{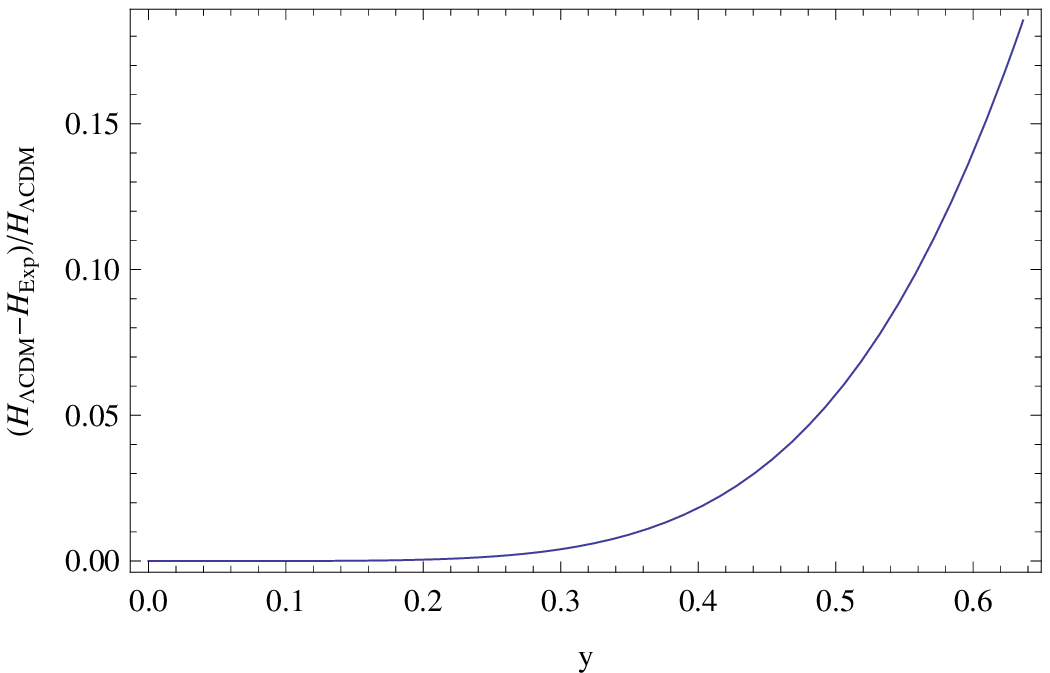}
\includegraphics[width=8.1cm]{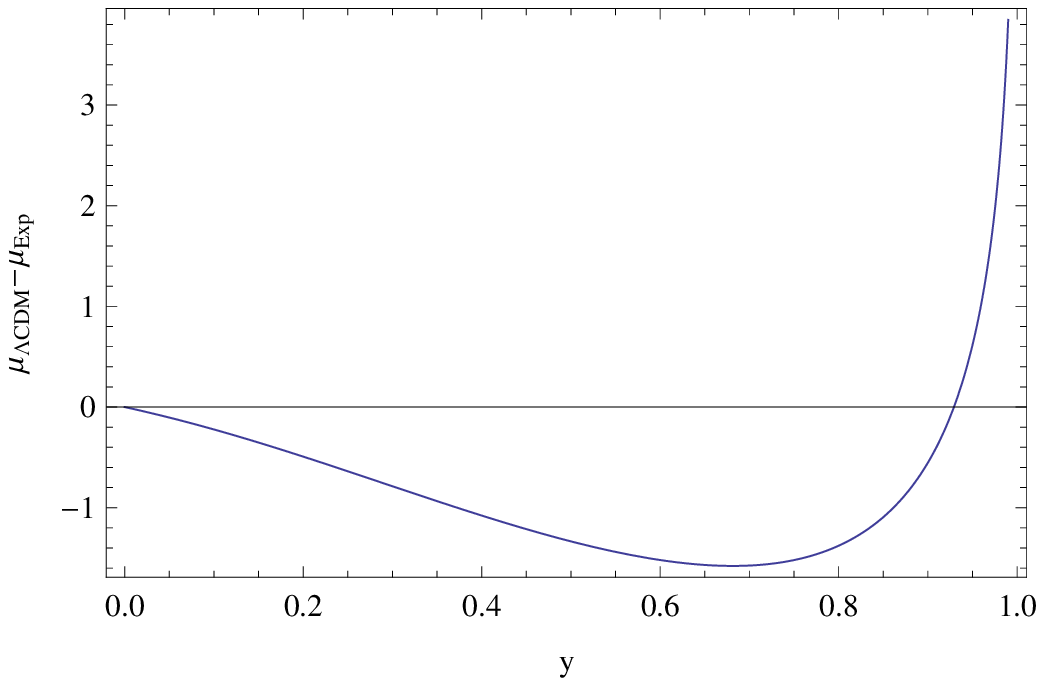}
\caption{The Hubble parameter departure and differences of distance modulus from $\Lambda$CDM model, where $\Omega_{m0}=0.27$ is fixed.}\label{fig:deviations}
\end{figure}

The Hubble parameter $H(z)$ can be expanded as
\begin{equation}
H(z)=H_0+\left.\frac{dH}{dz}\right|_0
z+\frac{1}{2}\left.\frac{d^2H}{dz^2}\right|_0
z^2+\frac{1}{3!}\left.\frac{d^3H}{dz^3}\right|_0 z^3+...,
\end{equation}
where the subscript '$0$' denotes the value at the present epoch and
$z=1/a(t)-1$. Via the relation
\begin{equation}
\frac{dt}{dz}=-\frac{1}{(1+z)H(z)},
\end{equation}
one has
\begin{eqnarray}
\left.\frac{dH}{dz}\right|_0&=&-\left.\frac{\dot{H}}{(1+z)H}\right|_0=(1+q_0)H_0,\\
\left.\frac{d^2H}{dz^2}\right|_0&=&\left.\frac{\ddot{H}}{(1+z)^2H^2}\right|_0+\left.\dot{H}\left(\frac{1}{(1+z)^2H}-\frac{\dot{H}}{(1+z)^2H^3}\right)\right|_0\nonumber\\
&=&(j_0+3q_0+2)H_0-(q_0^2+3q_0+2)H_0\nonumber\\
&=&(j_0-q_0^2)H_0,\\
\left.\frac{d^3H}{dz^3}\right|_0&=&-\left.\frac{H^{(3)}}{(1+z)^3H^3}\right|_0-\left.3\frac{\ddot{H}}{(1+z)^2H^2}\left(\frac{1}{1+z}+\frac{1}{H}\frac{dH}{dz}\right)\right|_0\nonumber\\
&+&\left.\frac{\dot{H}}{(1+z)H}\left[-\frac{2}{(1+z)^2}
-\frac{2}{(1+z)H}\frac{dH}{dz}-\frac{2}{H^2}(\frac{dH}{dz})^2+\frac{1}{H}\frac{d^2H}{dz^2}\right]\right|_0\nonumber\\
&=&(6+12q_0+3q_0^2+4j_0-s_0)H_0-3(2+3q_0+j_0)(2+q_0)H_0\nonumber\\
&+&(1+q_0)\left[2+2(1+q_0)+2(1+q_0)^2+q_0^2-j_0\right]H_0\nonumber\\
&=&[3q_0^3+3q_0^2-j_0(3+4q_0)-s_0]H_0,
\end{eqnarray}
where the cosmographic parameters are defined as follows
\begin{eqnarray}
H_0&\equiv&\left.\frac{da(t)}{dt}\frac{1}{a(t)}\right|_0\equiv\left.\frac{\dot{a}(t)}{a(t)}\right|_0,\\
q_0&\equiv&-\left.\frac{1}{H^2}\frac{d^2a(t)}{dt^2}\frac{1}{a(t)}\right|_0\equiv-\left.\frac{1}{H^2}\frac{\ddot{a}(t)}{a(t)}\right|_0,\label{eq:q_0}\\
j_0&\equiv&\left.\frac{1}{H^3}\frac{d^3a(t)}{dt^3}\frac{1}{a(t)}\right|_0\equiv\left.\frac{1}{H^3}\frac{a^{(3)}(t)}{a(t)}\right|_0,\label{eq:j_0}\\
s_0&\equiv&\left.\frac{1}{H^4}\frac{d^4a(t)}{dt^4}\frac{1}{a(t)}\right|_0\equiv\left.\frac{1}{H^4}\frac{a^{(4)}(t)}{a(t)}\right|_0.\label{eq:s_0}\\
\end{eqnarray}
Then the Hubble parameter can be rewritten in terms of the
cosmographic parameters as
\begin{equation}
H(z)=H_0\left\{1+(1+q_0)z+(j_0-q_0^2)z^2/2+[3q_0^3+3q_0^2-j_0(3+4q_0)-s_0]z^3/6+...\right\}.
\end{equation}
For a spatially flat FRW universe, the luminosity distance can also
be expanded in terms of redshift $z$ with the cosmographic
parameters
\begin{eqnarray}
d_L(z)&=&cH_0^{-1}\left\{z+(1-q_0)z^2/2-\left(1-q_0-3q_0^2+j_0\right)z^3/6\right.\nonumber \\ && \nonumber \\
&+&\left.\left[2-2q_0-15q_0^2-15q_0^3+5j_0+10q_0j_0+
s_0\right]z^4/24+...\right\}.
\end{eqnarray}
Via the relation $d_A(z)=d_L(z)/(1+z)^2$, one has the expansion of
$d_A(z)$
\begin{eqnarray}
d_A(z)&=&cH_0^{-1}\left\{z-(3+q_0)z^2/2+(11-j_0+7q_0+3q_0^2)z^3/6\right.\nonumber\\
&+&
\left.(-50+13j_0-46q_0+10j_0q_0-39q_0^2-15q_0^3+s_0)z^4/24+...\right\}.
\end{eqnarray}
To avoid problems with the convergence of the series for the highest
redshift objects, these relations are recast in terms of the new
variable $y=z/(1+z)$ \cite{ref:ztoy,ref:ztoy2}
\begin{eqnarray}
H(y)&=&H_0\left\{1+(1+q_0)y+(1+q_0+j_0/2-q_0^2/2)y^2+\left(6+3j_0+6q_0-4q_0j_0-3q_0^2+3q_0^3-s_0\right)y^3/6\right.\nonumber\\
&+&\left.\left(1+q_0-2j_0q_0+3q_0^3/2-s_0/2\right) y^4+\mathcal{O}(y^5)\right\}\\
d_L(y)&=&cH_0^{-1}\left\{y+(3-q_0)y^2/2+(11-j_0-5q_0+3q_0^2)y^3/6\right.\nonumber\\
&+&\left.\left(50-7j_0-26q_0+10q_0j_0+21q_0^2-15q_0^3+s_0\right)y^4/24+\mathcal{O}(y^5)\right\}\\
d_A(y)&=&cH_0^{-1}\left\{y-(1+q_0)y^2/2-(1+j_0-q_0-3q_0^2)y^3/6\right.\nonumber\\
&+&\left.(-2+j_0+2q_0+10j_0q_0-3q_0^2-15q_0^3+s_0)y^4/24+\mathcal{O}(y^5)\right\}
\end{eqnarray}
With this new variable, $z\in (0,\infty)$ is mapped into
$y\in(0,1)$. And the right behavior for series convergence at any
distance can be retrieved in principle \cite{ref:ztoy,ref:ztoy2}. 
When the convergence problem is solved, one has to concern the expansion truncation issue. Of course, with higher orders expansion, more accurate approximation would be obtained. However, in this way, one has to introduce more model parameters beyond $H_0$, $q_0$, $j_0$ and $s_0$. How to keep the balance between the free model parameters (or expansion truncation) and comic observational data points is another complicated problem. That is beyond the scope of this paper. But we'd like to point out that the way out may be the so-call Bayesian evidence method. In fact, we can show the deviations of the expansions from $\Lambda$CDM model. For illustration, with fixed value of $\Omega_{m0}=0.27$,
the relative departure of Hubble parameter from $\Lambda$CDM model (the left panel) and differences of distance modulus to $\Lambda$CDM model (the right panel) are shown in Fig. \ref{fig:deviations}. Actually, in the redshift range ($z\in [0,1.75]$, please see Tab. \ref{Hubbledata}) of the observational Hubble parameters, the relative departure of $\Lambda$CDM model is up to $\sim 20\%$ which is almost the same of order of error bars of OHD. In the right panel of Fig. \ref{fig:deviations}, the difference of distance modulus between the expansion of luminosity distance and that of $\Lambda$CDM model is shown. At high redshift $y\sim 1$, the departure is larger up to $4$. In the redshift range of this paper, $z\in[0,9]$, the difference of distance modulus is less than $1.6$. So, up to the fourth oder of $y$, these expansions are safe.

As the reader has noticed the Taylor expansion is up to snap parameter
$s_0$, with these cosmographic parameters the Hubble parameter is of
the order $z^3$. However, $d_L(z)$ and $d_A(z)$ are of the order
$z^4$. This is really from the fact that the Hubble parameter has
contained one order derivative of time $t$. When it is up to the
same order of $d_L(z)$ and $d_A(z)$, an extra new parameter has to
be introduced. So we will classify the data sets on hand into two
cases with (Case I: SN+BAO+GRBs) or without (Case II:
SN+BAO+GRBs+OHD) the observational Hubble data. Another reason is
that the cosmic observational data sets of SN and GRBs do not have
constraint to Hubble parameter $H_0$. That can be seen clearly from
the left panel of Fig. \ref{fig:cases} in this paper. So, to fix the
current value of Hubble parameter, the OHD data sets should be
added. The reader can also see that the BAO data set is helpful to
shrink the parameter space.

\section{Results and Discussion} \label{sec:results}

In our calculations, we have taken the total likelihood function
$L\propto e^{-\chi^2/2}$ to be the products of the separate
likelihoods of SN (with systematic errors), BAO, GRBs and OHD. Then
we get $\chi^2$
\begin{eqnarray}
\chi^2=\chi^2_{SN}+\chi^2_{BAO}+\chi^2_{GRBs}+\chi^2_{OHD},
\end{eqnarray}
where the separate likelihoods of SN, BAO, GRBs, OHD and the current
observational data sets used in this paper are shown in the Appendix
\ref{app:obervations}.

\begin{figure}[!htbp]
\includegraphics[width=8.2cm]{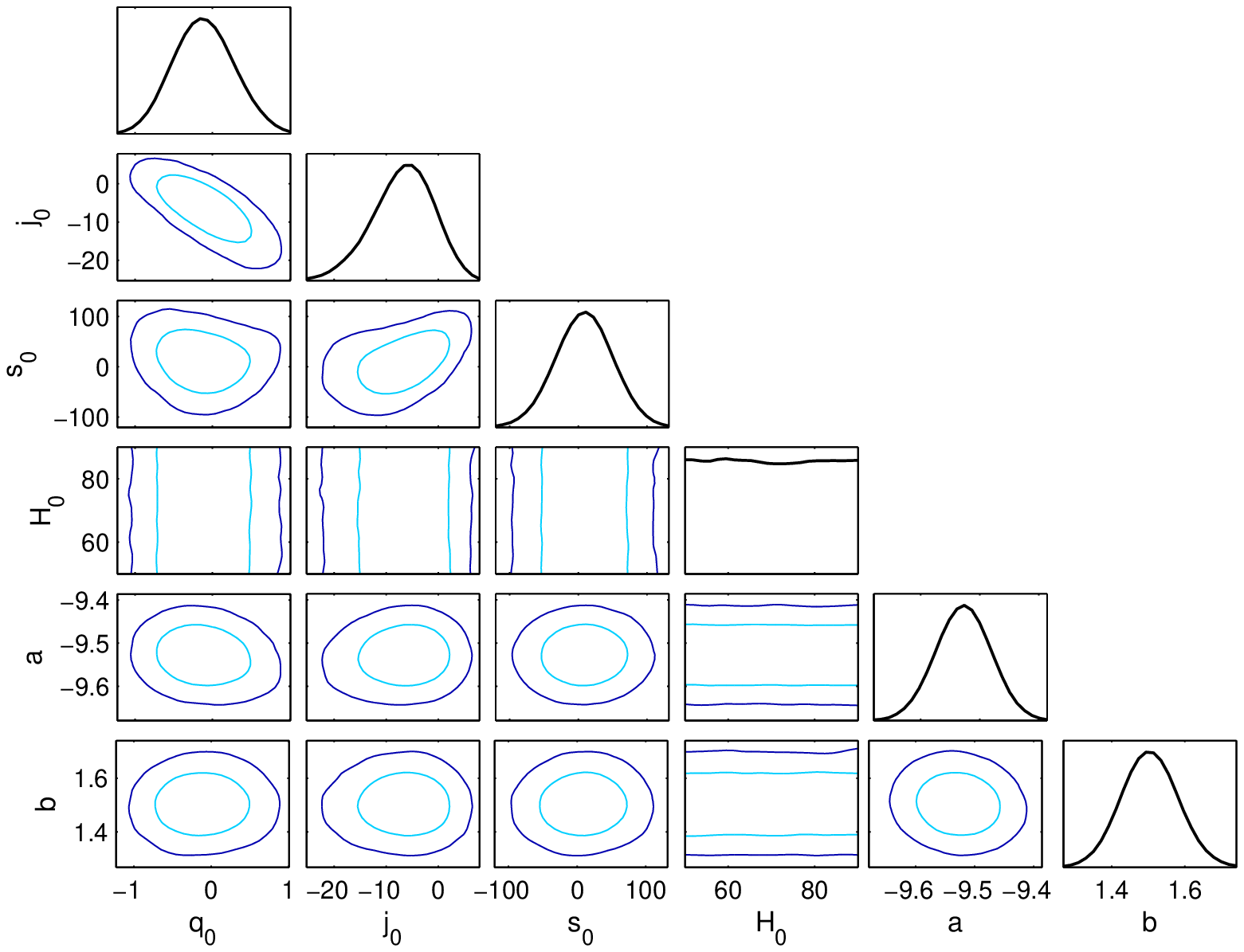}
\includegraphics[width=8.2cm]{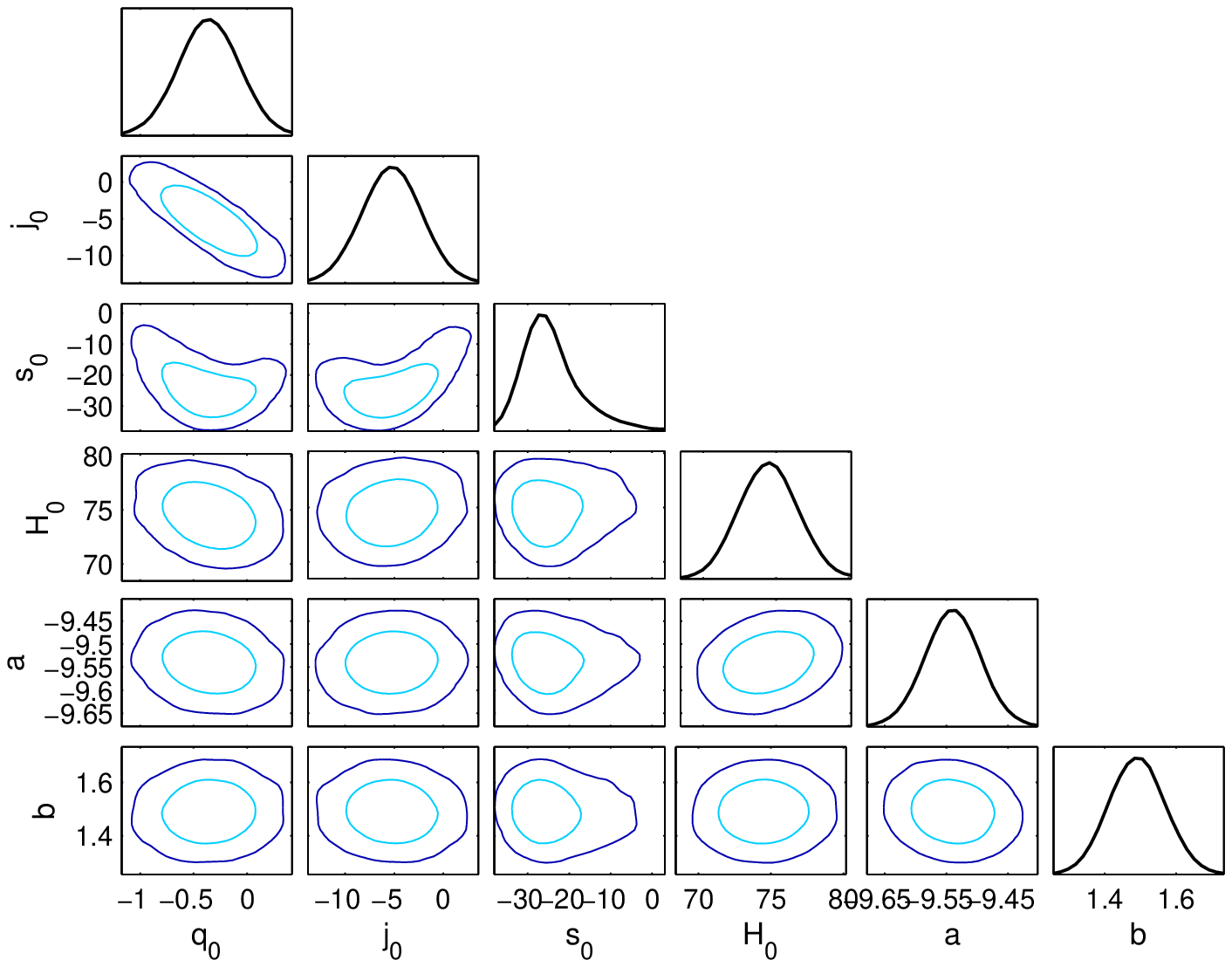}
\caption{The 1-D marginalized distribution and 2-D contours of model
parameter spaces with $1\sigma$, $2\sigma$ regions. Left Panel: Case
I: SN+BAO+GRBs. Right Panel: Case II:
SN+BAO+GRBs+OHD.}\label{fig:cases}
\end{figure}

\begin{figure}[!htbp]
\includegraphics[width=8.2cm]{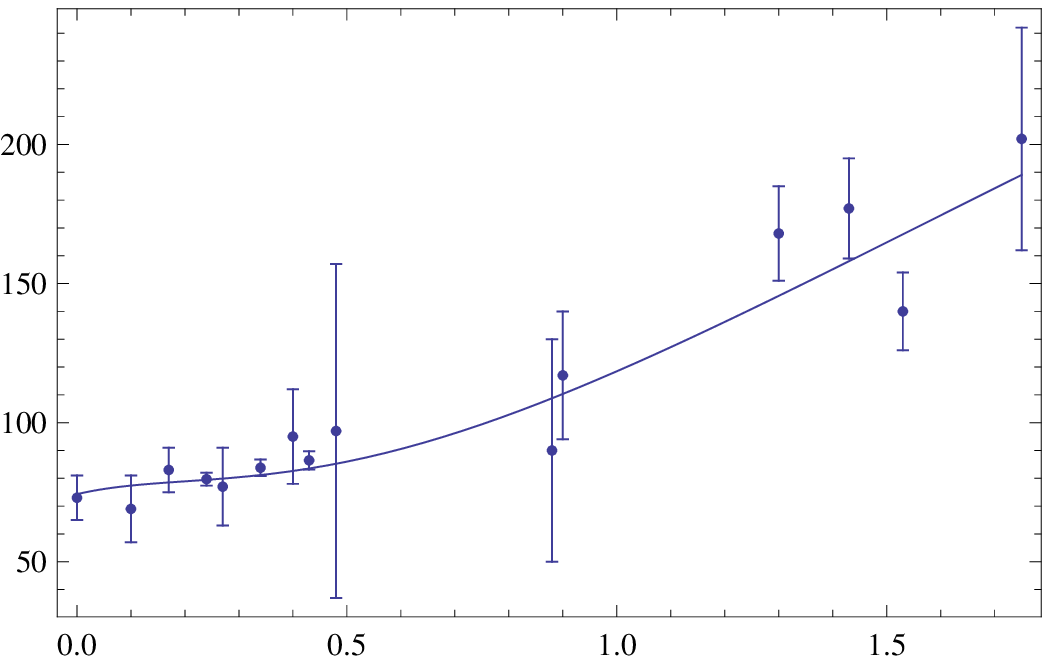}
\includegraphics[width=8.1cm]{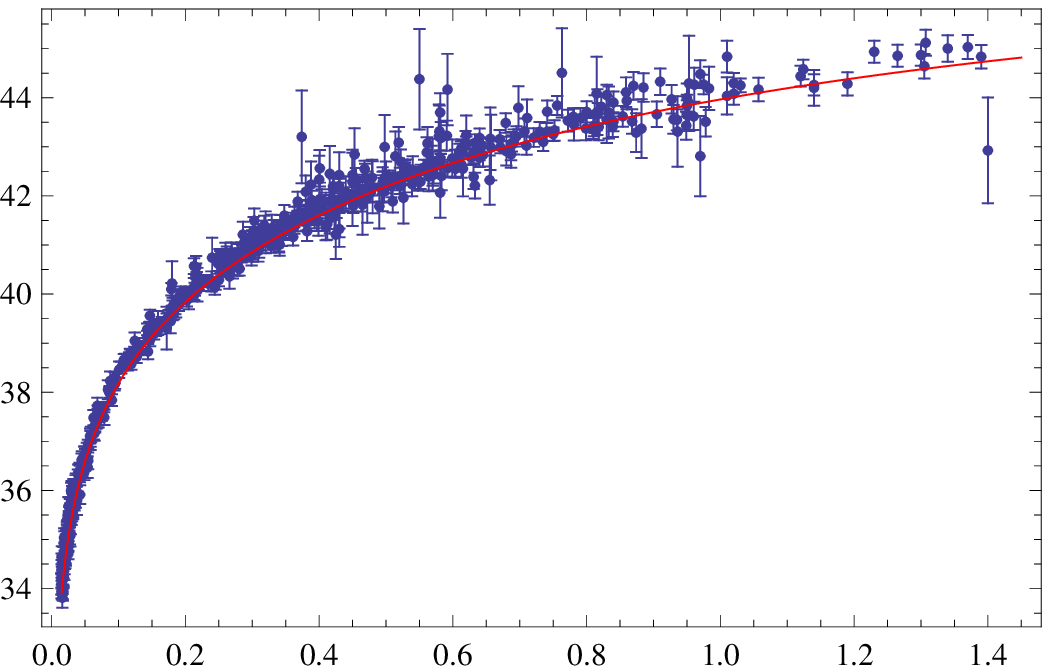}
\caption{The Hubble parameter and distance modulus with respect to redshift $z$, where the best fitted model parameter values in the third row of Tab. \ref{tab:results} are adopted. }\label{fig:hubmod}
\end{figure}

In our analysis, we perform a global fitting to determine the
cosmographic parameters using the MCMC method. Our code is based on
the publicly available {\bf CosmoMC} package \cite{ref:MCMC}. The
results are shown in Table \ref{tab:results} and Figure \ref{fig:cases}.
\begin{table}
\begin{center}
\begin{tabular}{cc|    cc|     cc|     cc|    cc|  cc |  cc | cc}
\hline\hline Model & & $\chi^2_{min}/d.o.f$ & & $H_0$ & & $q_0$ & &
$j_0$& & $s_0$ & & $a$ & & $b$
\\ \hline
Case I && $656.821/661$ && $-$ & & $-0.150^{+0.887}_{-0.752}$ &&
$-5.848^{+10.0999}_{-14.412}$ && $-81.268^{+91.708}_{-88.218}$ &
& $-9.522^{+0.0909}_{-0.104}$ & & $1.499^{+0.173}_{-0.159}$\\
\hline Case II && $670.954/676$ && $74.299^{+4.932}_{-4.287}$ & &
$-0.386^{+0.655}_{-0.618}$ && $-4.925^{+6.658}_{-7.297}$ &&
$-26.404^{+20.964}_{-9.097}$ &
& $-9.540^{+0.104}_{-0.0999}$ & & $1.483^{+0.187}_{-0.166}$\\
\hline\hline
\end{tabular}
\caption{The results of $\chi^2_{min}$, $H_0$, $q_0$, $j_0$ and
$s_0$ in Case I (SN+BAO+GRBs) and Case II (SN+BAO+GRBs+OHD), where
{\rm d.o.f} denotes the degree of freedom. $a$ and $b$ are
parameters from Amati's correlation of GRBs, for their definition please see Eq. \ref{eq:calib}.}\label{tab:results}
\end{center}
\end{table}
And the evolution curves of the Hubble parameter and distance modulus with respect to redshift $z$ are shown in Fig. \ref{fig:hubmod} where the best fitted values of model parameters are adopted from the third row of Tab. \ref{tab:results}.

One can clearly see that when the observational Hubble data are used
the $1\sigma$ error parameters space is shrunk remarkably. Put in
other words, the figure of merit is improved tremendously. It is
really from the fact that the Hubble parameter $H$ is expressed in
terms of $z$ or $y$ with combined cosmographic parameters
coefficients. Also, from the second row of Table \ref{tab:results},
one has noticed that the BAO data set is helpful to break the
degeneracy and shrink the parameter space. We can test the
reliability by comparing the result with spatially flat $\Lambda$CDM
model. For the spatially flat $\Lambda$CDM model, we can easily find
the corresponding deceleration, jerk and snap parameters
respectively
\begin{eqnarray}
q_0&=&\frac{3}{2}\Omega_{m0}-1,\\
j_0&=&1,\\
s_0&=&1-\frac{9}{2}\Omega_{m0}.
\end{eqnarray}
When $\Omega_{m0}$ varies in the range $\Omega_{m0}\in[0,1]$, $q_0$
and $s_0$ will be in the ranges $q_0\in[-1,0.5]$ and $s_0\in
[-3.5,1]$ respectively. For comparing the best fit values of
cosmographic parameters in Case II with the spatially flat
$\Lambda$CDM model, where the same data sets combination is used to
constrain the flat $\Lambda$CDM model, one finds the corresponding
result: $\Omega_{m0}=0.270_{-0.0355}^{+0.0403}$,
$a=-9.398_{-0.0723}^{+0.0708}$ and $b=1.602_{-0.128}^{+0.135}$. One
can clearly see that for the best fit value of $\Omega_{m0}=0.270$
in flat $\Lambda$CDM model the derived $q_0=-0.595$ and $j_0=1$ are
consistent with the results obtained from cosmographic approach in
$1\sigma$ region. However, the value of $s_0=-0.215$ of flat
$\Lambda$CDM model is out the range of the $1\sigma$ region of
cosmographic approach. As discussed in Ref. \cite{ref:kinematicXu},
once the parameterized deceleration parameter $q(z)=q_0+q_1z/(1+z)$
\cite{ref:xupq} is known, one can find the relation
$q_1=-q_0-2q_0^2-j_0$. Also one can find other interesting
relations, for example the relations between the modified gravity
theory, DGP brane world model, $w=constant$, CPL parameterized
equation of state of dark energy \cite{ref:ztoy2} and cosmographic
parameters were investigated in Ref. \cite{ref:cosmoDai}, see also
in Ref. \cite{ref:kinematicXu}.

\section{Conclusion}\label{sec:con}

In this paper, the cosmographic approach is reconsidered by using
cosmic observational data which include SN Union2, BAO, GRBs and OHD
via MCMC method. We find the best fit values of cosmographic
parameters in $1\sigma$ ranges: $H_0=74.299^{+4.932}_{-4.287}$,
$q_0=-0.386^{+0.655}_{-0.618}$, $j_0=-4.925^{+6.658}_{-7.297}$ and
$s_0=-26.404^{+20.964}_{-9.097}$ which are improved remarkably.
Comparing with the spatially flat $\Lambda$CDM model, one can find
out that the derived values of $q_0$ and $j_0$ in flat $\Lambda$CDM
are consistent with the results obtained from cosmographic approach
in $1\sigma$ region. But the value of $s_0$ of flat $\Lambda$CDM
model is out of the $1\sigma$ region of cosmographic best fit value.
As investigated, the BAO data set is helpful to shrink the parameter
space. When the OHD data sets are added, the parameters space is
improved remarkably. The reason is from the fact that the Hubble
parameter $H$ is expressed in terms of $z$ or $y$ with combined
cosmographic parameters coefficients. In summary, the main points of this paper are that 1). BAO and OHD are are helpful to shrink the parameter space. 2). The calibration of GRBs and constraint to cosmographic parameters are carried out synchronously. It is away from the so-called circular problem. 

\acknowledgements{This work is supported by NSF (10703001) and the
Fundamental Research Funds for the Central Universities (DUT10LK31).
We thank Dr. V. Vitagliano for his correspondence and anonymous referee for the constructive and helpful comments.}

\appendix

\section{Cosmic Observational Data Sets}\label{app:obervations}

\subsection{Type Ia Supernovae}

Recently, SCP (Supernova Cosmology Project) collaboration released
their Union2 dataset which consists of 557 SN Ia \cite{ref:SN557}.
The distance modulus $\mu(z)$ is defined as
\begin{equation}
\mu_{th}(z)=5\log_{10}[\bar{d}_{L}(z)]+\mu_{0},
\end{equation}
where $\bar{d}_L(z)$ is the Hubble-free luminosity distance $H_0
d_L(z)/c=H_0 d_A(z)(1+z)^2/c$, with $H_0$ the Hubble constant, and
$\mu_0\equiv42.38-5\log_{10}h$ through the re-normalized quantity
$h$ as $H_0=100 h~{\rm km ~s}^{-1} {\rm Mpc}^{-1}$. Where $d_L(z)$
is defined as
\begin{equation}
d_L(z)=(1+z)r(z),\quad r(z)=\frac{c}{H_0\sqrt{|\Omega_{k}|}}{\rm
sinn}\left[\sqrt{|\Omega_{k}|}\int^z_0\frac{dz'}{E(z')}\right]
\end{equation}
where $E^2(z)=H^2(z)/H^2_0$. Additionally, the observed distance
moduli $\mu_{obs}(z_i)$ of SN Ia at $z_i$ are
\begin{equation}
\mu_{obs}(z_i) = m_{obs}(z_i)-M,
\end{equation}
where $M$ is their absolute magnitudes.

For the SN Ia dataset, the best fit values of the parameters $p_s$
can be determined by a likelihood analysis, based on the calculation
of
\begin{eqnarray}
\chi^2(p_s,M^{\prime})&\equiv& \sum_{SN}\frac{\left\{
\mu_{obs}(z_i)-\mu_{th}(p_s,z_i)\right\}^2} {\sigma_i^2}  \nonumber\\
&=&\sum_{SN}\frac{\left\{ 5 \log_{10}[\bar{d}_L(p_s,z_i)] -
m_{obs}(z_i) + M^{\prime} \right\}^2} {\sigma_i^2}, \label{eq:chi2}
\end{eqnarray}
where $M^{\prime}\equiv\mu_0+M$ is a nuisance parameter which
includes the absolute magnitude and the parameter $h$. The nuisance
  parameter $M^{\prime}$ can be marginalized over
analytically \cite{ref:SNchi2} as
\begin{equation}
\bar{\chi}^2(p_s) = -2 \ln \int_{-\infty}^{+\infty}\exp \left[
-\frac{1}{2} \chi^2(p_s,M^{\prime}) \right] dM^{\prime},\nonumber
\label{eq:chi2marg}
\end{equation}
resulting to
\begin{equation}
\bar{\chi}^2 =  A - \frac{B^2}{C} + \ln \left( \frac{C}{2\pi}\right)
, \label{eq:chi2mar}
\end{equation}
with
\begin{eqnarray}
A&=&\sum_{i,j}^{SN}\left\{5\log_{10}
[\bar{d}_L(p_s,z_i)]-m_{obs}(z_i)\right\}\cdot {\rm
Cov}^{-1}_{ij}\cdot \left\{5\log_{10}
[\bar{d}_L(p_s,z_j)]-m_{obs}(z_j)\right\},\nonumber\\
B&=&\sum_i^{SN} {\rm Cov}^{-1}_{ij}\cdot \left\{5\log_{10}
[\bar{d}_L(p_s,z_j)]-m_{obs}(z_j)\right\},\nonumber \\
C&=&\sum_i^{SN} {\rm Cov}^{-1}_{ii},\label{eq:SNsyserror}
\end{eqnarray}
where ${\rm Cov}^{-1}_{ij}$ is the inverse of covariance matrix with
or without systematic errors. One can find the details in Ref.
\cite{ref:SN557} and the web site
\footnote{http://supernova.lbl.gov/Union/} where the covariance
matrix with or without systematic errors are included. Relation
(\ref{eq:chi2}) has a minimum at the nuisance parameter value
$M^{\prime}=B/C$, which contains information of the values of $h$
and $M$. Therefore, one can extract the values of $h$ and $M$
provided the knowledge of one of them. Finally, the expression
\begin{equation}
\chi^2_{SN}(p_s,B/C)=A-(B^2/C),\label{eq:chi2SN}
\end{equation}
which coincides to Eq. (\ref{eq:chi2mar}) up to a constant, is often
used in the likelihood analysis \cite{ref:smallomega,ref:SNchi2}.
Thus in this case the results will not be affected by a flat
$M^{\prime}$ distribution. It worths noting that the results will be
different with or without the systematic errors. In this work, all
results are obtained with systematic errors.

\subsection{BAO}
The BAO are detected in the clustering of the combined 2dFGRS and
SDSS main galaxy samples, and measure the distance-redshift relation
at $z = 0.2$. BAO in the clustering of the SDSS luminous red
galaxies measure the distance-redshift relation at $z = 0.35$. The
observed scale of the BAO calculated from these samples and from the
combined sample are jointly analyzed using estimates of the
correlated errors, to constrain the form of the distance measure
$D_V(z)$ \cite{ref:Okumura2007,ref:Eisenstein05,ref:Percival}
\begin{equation}
D_V(z)=\left[(1+z)^2 D^2_A(z) \frac{cz}{H(z)}\right]^{1/3},
\label{eq:DV}
\end{equation}
where $D_A(z)$ is the proper (not comoving) angular diameter
distance which has the following relation with $d_{L}(z)$
\begin{equation}
D_A(z)=\frac{d_{L}(z)}{(1+z)^2}.
\end{equation}
Matching the BAO to have the same measured scale at all redshifts
then gives \cite{bao:dvratio}
\begin{equation}
D_{V}(0.35)/D_{V}(0.2)=1.736\pm0.065.
\end{equation}
Then, the $\chi^2_{BAO}(p_s)$ is given as
\begin{equation}
\chi^2_{BAO}(p_s)=\frac{\left[D_{V}(0.35)/D_{V}(0.2)-1.736\right]^2}{0.065^2}\label{eq:chi2BAO}.
\end{equation}

\subsection{Gamma Ray Bursts}

Following \cite{ref:Schaefer}, we consider the well-known Amati's
$E_{p,i}-E_{iso}$ correlation \cite{ref:Amati'srelation,r16,r17,r18}
in GRBs, where $E_{p,i}=E_{p,obs}(1+z)$ is the cosmological
rest-frame spectral peak energy, and $E_{iso}$ is the isotropic
energy
\begin{equation}
E_{iso}=4\pi d^2_LS_{bolo}/(1+z)
\end{equation}
in which $d_L$ and $S_{bolo}$ are the luminosity distance and the
bolometric fluence of the GRBs respectively. Following
\cite{ref:Schaefer}, we rewrite the Amati's relation as
\begin{equation}
\log\frac{E_{iso}}{{\rm erg}}=a+b\log\frac{E_{p,i}}{300{\rm
keV}}.\label{eq:calib}
\end{equation}

In \cite{ref:cosmographygrb}, the correlation parameters were
calibrated via cosmographic approach. Following this method, we take
correlation parameters $a$ and $b$ as free parameters when GRBs is
used as a cosmic constraint. We fit the Amati's relation through the
minimization $\chi^2$ given by \cite{ref:Schaefer}
\begin{equation}
\chi^2_{GRBs}(p_s)=\sum^N_{i=1}\frac{y_i-a-bx_i}{\sigma^2_{y,i}+b^2\sigma^2_{x,i}+\sigma^2_{sys}},
\end{equation}
where
\begin{eqnarray}
x_i&=&\log\frac{E_{p,i}}{300{\rm keV}}\\
y_i&=&\log\frac{E_{iso}}{{\rm erg}}=\log\frac{4\pi
S_{bolo,i}}{1+z}+2\log\bar{d}_L
\end{eqnarray}
where $\bar{d}_L$ is defined as \cite{ref:wang}
\begin{equation}
\bar{d}_L=H_0(1+z)r(z)/c,
\end{equation}
and the errors are calculated by using the error propagation law
\cite{ref:errors}:
\begin{eqnarray}
\sigma_{x,i}&=&\frac{\sigma_{E_{p,i}}}{\ln10E_{p,i}}\\
\sigma_{y,i}&=&\frac{\sigma_{S_{bolo,i}}}{\ln10S_{bolo,i}}.
\end{eqnarray}
Here $N=109$ GRBs data points are taken from \cite{ref:Wei109}. The
$\chi^2$ is large and dominated by the systematic errors, and the
statistical errors on $a$ and $b$ are small. In general the
systematic error $\sigma_{sys}$ can be derived by required
$\chi^2=\nu$ (the degrees of freedom) \cite{ref:Schaefer}. Here, we
take the value of $\sigma^2_{sys}=0.324$ from Table 1. of the case
of $\Omega_{m0}=0.27$ in Ref. \cite{ref:GRBsXu}. In fact, the
concrete value does affect the results concluded in this paper. At
last, the total error is
$\sigma^2_{tot}=\sigma^2_{stat}+\sigma^2_{sys}$. It would be noticed
that in our case, the best fit value of $a$ will be less than
$2\log(c/H_0)$ in the definition of luminosity distance
$d_L=(1+z)r(z)$ \cite{ref:wang}.

\subsection{Observational Hubble Data}

The observational Hubble data are based on differential ages of the
galaxies \cite{ref:JL2002}. In \cite{ref:JVS2003}, Jimenez {\it et
al.} obtained an independent estimate for the Hubble parameter using
the method developed in \cite{ref:JL2002}, and used it to constrain
the EOS of dark energy. The Hubble parameter depending on the
differential ages as a function of redshift $z$ can be written in
the form of
\begin{equation}
H(z)=-\frac{1}{1+z}\frac{dz}{dt}.
\end{equation}
So, once $dz/dt$ is known, $H(z)$ is obtained directly
\cite{ref:SVJ2005}. By using the differential ages of
passively-evolving galaxies from the Gemini Deep Deep Survey (GDDS)
\cite{ref:GDDS} and archival data
\cite{ref:archive1,ref:archive2,ref:archive3,ref:archive4,ref:archive5,ref:archive6},
Simon {\it et al.} obtained $H(z)$ in the range of $0.1\lesssim z
\lesssim 1.8$ \cite{ref:SVJ2005}. In \cite{ref:0907}, Stern {\it et
al.} used the new data of the differential ages of
passively-evolving galaxies at $0.35<z<1$ from Keck observations,
SPICES survey and VVDS survey. The twelve observational Hubble data
from \cite{ref:0905,ref:0907,ref:SVJ2005} are list in Table
\ref{Hubbledata}. Here, we use the value of Hubble constant
$H_0=74.2\pm3.6 {\rm km ~s}^{-1} {\rm Mpc}^{-1}$, which is obtained
by observing 240 long-period Cepheids in \cite{ref:0905}. As pointed
out in \cite{ref:0905}, the systematic uncertainties have been
greatly reduced by the unprecedented homogeneity in the periods and
metallicity of these Cepheids. For all Cepheids, the same instrument
and filters are used to reduce the systematic uncertainty related to
flux calibration.
\begin{table}[htbp]
\begin{center}
\begin{tabular}{c|llllllllllll}
\hline\hline
 $z$ &\ 0 & 0.1 & 0.17 & 0.27 & 0.4 & 0.48 & 0.88 & 0.9 & 1.30 & 1.43 & 1.53 & 1.75  \\ \hline
 $H(z)\ ({\rm km~s^{-1}\,Mpc^{-1})}$ &\ 74.2 & 69 & 83 & 77 & 95 & 97 & 90 & 117 & 168 & 177 & 140 & 202  \\ \hline
 $1 \sigma$ uncertainty &\ $\pm 3.6$ & $\pm 12$ & $\pm 8$ & $\pm 14$ & $\pm 17$ & $\pm 60$ & $\pm 40$
 & $\pm 23$ & $\pm 17$ & $\pm 18$ & $\pm 14$ & $\pm 40$ \\
\hline
\end{tabular}
\end{center}
\caption{\label{Hubbledata} The observational $H(z)$
data~\cite{ref:0905,ref:0907}.}
\end{table}
In addition, in \cite{ref:0807}, the authors took the BAO scale as a
standard ruler in the radial direction, called "Peak Method",
obtaining three more additional data: $H(z=0.24)=79.69\pm2.32,
H(z=0.34)=83.8\pm2.96,$ and $H(z=0.43)=86.45\pm3.27$, which are
model and scale independent. Here, we just consider the statistical
errors.

The best fit values of the model parameters are determined by
minimizing
\begin{equation}
\chi_{OHD}^2(p_s)=\sum_{i=1}^{15} \frac{[H_{th}(p_s;z_i)-H_{
obs}(z_i)]^2}{\sigma^2(z_i)},\label{eq:chi2H}
\end{equation}
where $p_s$ denotes the parameters contained in the model, $H_{th}$
is the predicted value for the Hubble parameter, $H_{obs}$ is the
observed value, $\sigma(z_i)$ is the standard deviation measurement
uncertainty, and the summation is over the $15$ observational Hubble
data points at redshifts $z_i$. The OHD was firstly used to
constrain cosmological model in \cite{ref:ZhangTJOHD}.

\end{document}